\documentclass[11pt,a4paper]{article}
\pdfoutput=1
\usepackage{jheppub}
\usepackage{rotating}
\usepackage{tikz}
\usepackage{tikz-feynman}
\usepackage{wrapfig}
\usepackage{amsfonts}
\usepackage{amsmath}
\usepackage{slashed}
\usepackage{amssymb}
\usepackage{latexsym}
\usepackage{multirow}
\usepackage{hyperref}
\usepackage{enumitem}
\usepackage{orcidlink}
\usepackage[makeroom]{cancel}
\usepackage{comment}
\hypersetup{colorlinks=true,citecolor=red,linkcolor=magenta,urlcolor=blue}
\usepackage[caption=false]{subfig}
\usepackage{relsize}
\usepackage{booktabs}
\usepackage{cleveref}
\usepackage{fancyhdr}
\usepackage{float}
\usepackage{graphicx}
\usepackage{microtype}
\usepackage{tabularx}
\usepackage[bottom]{footmisc}

\definecolor{ForestGreen}{RGB}{34,139,34}
\newcommand{\ce}[1]{{\color{red}{{#1}}}}

\newcommand{\hsd}{\zeta}

\newcommand{\lag}{\mathcal{L}}
\newcommand{\boxh}{\Box h}
\newcommand{\abb}{a_{\Box \Box}}
\newcommand{\obb}{\mathcal{O}_{\Box \Box}}
\newcommand{\obxbx}{\boxh \boxh}
\newcommand{\feynarts}{\texttt{FeynArts}~\cite{Hahn:2000kx}}
\newcommand{\formcalc}{\texttt{FormCalc}~\cite{Hahn:1998yk}}
\newcommand{\packagex}{\texttt{Package-X}~\cite{Patel:2016fam}}

\renewcommand{\square}{\Box}

\preprint{DESY-25-171, IPPP/25/76}
\title{Assessing (H)EFT theory errors by pitting EoM against Field Redefinitions}
\abstract{Truncations of effective field theory expansions are technically necessary but inherently intertwined with the redundancies of general field redefinitions. This can be viewed as a juxtaposition of power-counting and theoretical uncertainties, which seek to estimate neglected higher-dimensional interactions through approaches based on community consensus. One can then understand the invariance of physics under field redefinitions as a data-informed validation of different power-counting schemes, or as a means of assigning theoretical errors in comparison with algebraic, equation of motion-based replacements. Such an approach generalises widely accepted procedures for estimating theoretical uncertainties within the SM to non-renormalisable interactions. We perform a case study for a representative example in Higgs Effective Field theory, focusing on universal Higgs properties tensioned against process-dependent sensitivity expectations.}
\author[a]{Rodrigo Alonso\orcidlink{0000-0002-7197-8845},} 
\author[b]{Christoph Englert\orcidlink{0000-0003-2201-0667},} 
\author[c]{Wrishik Naskar\orcidlink{0000-0002-4357-8991},}
\author[a]{Shakeel Ur Rahaman\orcidlink{0000-0001-6153-8187}}
\affiliation[a]{Institute for Particle Physics Phenomenology, Department of Physics, Durham University,\\South Road, Durham DH1 3LE, United Kingdom}
\affiliation[b]{Department of Physics and Astronomy, University of Manchester, Oxford Road,\\Manchester M13 9PL, United Kingdom}
\affiliation[c]{Deutsches Elektronen-Synchrotron DESY, Notkestraße 85, 22607 Hamburg, Germany}
\emailAdd{rodrigo.alonso-de-pablo@durham.ac.uk}
\emailAdd{christoph.englert@manchester.ac.uk}
\emailAdd{wrishik.naskar@desy.de}
\emailAdd{shakeel.u.rahaman@durham.ac.uk}
\begin{document}
\maketitle
\flushbottom
\section{Introduction}
\label{sec:intro}
The lack of direct evidence for new physics beyond the Standard Model (BSM), especially at the high-energy frontier explored by the Large Hadron Collider (LHC), is puzzling given the apparent shortfalls of the SM to explain a range of experimentally observed phenomena, ranging from dark matter, over baryogenesis, to fine-tuning. The experimental observations at the LHC have put significant pressure on well-motivated candidate theories beyond the Standard Model. This has refocused phenomenological efforts toward model-independent techniques that present themselves through the application of, e.g., Standard Model Effective Field Theory (SMEFT)~\cite{Weinberg:1978kz,Grzadkowski:2010es,Brivio:2017vri,Aebischer:2025qhh}. A working assumption of SMEFT is that the Higgs mechanism is well-approximated by the Standard Model (SM). This is motivated by minimalism, addressing observed electroweak correlations, such as the smallness of the $\rho$ parameter. By construction, most theories that contain the SM as a low-energy limit can be approached via SMEFT, potentially requiring operators of higher dimension than six to match the expected experimental precision phenomenologically~\cite{Dawson:2021xei,Dawson:2022cmu,Ellis:2023zim,Dawson:2024ozw,Adhikary:2025gdh,Henning:2014wua, Drozd:2015rsp, Ellis:2016enq, delAguila:2016zcb, Ellis:2017jns, Kramer:2019fwz, 
Banerjee:2023iiv, Banerjee:2023xak, Chakrabortty:2023yke, Cohen:2020fcu, Dittmaier:2021fls}. 

It is worth highlighting, however, that the emergence of custodial isospin in the SM is an accidental consequence of the Higgs field's quantum numbers and the requirement of renormalisability. Similar to the baryon number as an accidental symmetry of the SM, it therefore deserves appropriate scrutiny. In other words, whilst electroweak physics seems to be surprisingly accurately modelled by an effective theory based on the SM, a more general vacuum structure of the SM's embedding into BSM theories~\cite{Alonso:2016btr,Alonso:2016oah,Cohen:2020xca} could indeed pinpoint phenomenological avenues towards BSM discovery that do not present themselves when assuming SMEFT~\cite{Alonso:2021rac,Englert:2023uug,Anisha:2024ryj,Grober:2025vse}.

A theoretical framework capable of accommodating such possibilities is Higgs Effective Field Theory (HEFT) \cite{Alonso:2023upf}. At the cost of an abundance of new parameters, HEFT offers a coarse-grained description of the electroweak vacuum structure, employing the Callan-Coleman-Wess-Zumino (CCWZ) formalism~\cite{Callan:1969sn,Coleman:1969sm} to address electroweak-scale phenomena. Within this approach, the physical Higgs boson emerges as a custodial singlet, treated independently from the Goldstone sector. Technically reminiscent of technicolour models extended by a techni-$\Sigma$ state, HEFT can consistently incorporate loop corrections for scenarios in which the Higgs boson arises only partially from an elementary sector, extending also to, e.g., pseudo-dilaton theories.

The link between the geometry of electroweak vacuum manifolds and their expected phenomenological results has been analysed in great detail in recent years~\cite{Alonso:2015fsp,Alonso:2016oah,Alonso:2022ffe,Alonso:2021rac,Alonso:2016btr}. This also includes EFT convergence of SMEFT-like theories from a HEFT perspective, highlighting field redefinitions as redundant parameters in a QFT. Here arises a practical problem very familiar from precision calculations in the SM itself. When viewed as a parametrising framework for SM deviations, SMEFT and HEFT (the latter with its various scales typically identified with the vacuum expectation value by convenience, see also Sec.~\ref{sec:pcpu}) introduce additional theoretical uncertainties by working at a given order in a characteristic expansion. In SMEFT, highlighting its relation to a UV theory containing the SM in the infrared, this is the inverse cut-off (or BSM mass scale) of the theory that is also used to construct bases of given dimensionality~\cite{Lehman:2015via, Lehman:2015coa, Henning:2015alf, Henning:2017fpj, Graf:2020yxt, Graf:2022rco, Sun:2022aag,Hanany:2010vu}. In HEFT, we typically consider loop-order expansions~\cite{Brivio:2016fzo,Buchalla:2013rka,Alonso:2012px,Sun:2022ssa}, or equivalently, an expansion in chiral dimension~\cite{Buchalla:2013eza,Buchalla:2017jlu}.

In the SM, there are motivated, yet arbitrary, choices regarding the specific parameterisation one can choose for the field operators and masses. Masses, in general, are highlighted via Nielsen identities~\cite{Nielsen:1975fs,Grassi:2001bz} as appropriate input parameters of a renormalisation programme when they can be observed (this excludes the light quark masses, which for practical purposes are chosen as zero in the parton model). On-shell renormalised canonical field operators are then an obvious choice, in particular as the on-shell scheme resums mass-related effects.\footnote{Of course, there are known pitfalls and the gauge-dependence of renormalised masses can provide technical obstacles. It is worthwhile mentioning that such technical issues are largely absent in HEFT, by construction.} Yet, any other choice of field could be considered, which would lead to an identical phenomenology when we supply the appropriate LSZ factor and sum over a sufficient number of loops in the theory. Given that this is technically not feasible, in the SM, this argument is usually reversed: The numerical impact of different renormalisation schemes (and changes of renormalisation scales for minimal-subtracted quantities) is considered a proxy for theoretical systematics.

When we consider this within effective, non-renormalisable extensions of the SM, we are drawn to the importance of power counting schemes that are sometimes adopted to establish hierarchies between different interactions~\cite{Gavela:2016bzc,Buchalla:2016sop}. A field redefinition, truncated for instance at a given order in the SMEFT expansion in a bottom-up approach, leads to different theoretical expectations. Yet again, purely on technical grounds, we are forced to consider a particular parametrisation in the expansion of new physics effects. 

Differences between parameterisations are only relevant when contrasted with the experimental precision available for an interpretation in the effective theory. It is not new that the precision of a reference measurement determines the accuracy of a higher-order programme of predicting correlations. However, when viewed in the context of EFTs, this is qualitatively different from SM interpretations. For the latter, input data are typically much more precisely known from processes other than those we seek to analyse. In the bottom-up EFT, uncertainties in EFT parameters are typically directly linked to the process that is used to obtain the QFT input itself (see, e.g.,~\cite{Anisha:2022ctm}).

It is the purpose of this work to clarify such uncertainties in the Higgs sector. Especially when viewed from a HEFT perspective, this enables an experimentally-driven approach to understanding theoretical limitations, not based on theoretical prejudice, but on the `quality' of the observed data or measurement. This work is organised as follows: In Sec.~\ref{sec:theory}, we discuss with a concrete example how different parameterisations of Lagrangians lead to (non)identical amplitude predictions. Particular emphasis is given to manipulations of Lagrangians using perturbative field redefinitions and equation-of-motion (EoM) relations. Their use is ubiquitous in the literature, but known to lead to physically different results at higher order~\cite{Criado:2018sdb}. Such differences can then be understood as a new source of theoretical uncertainty relevant for the experimental EFT interpretation at present and future colliders. We perform a representative case study for a specific Higgs boson-related HEFT interaction in Sec.~\ref{sec:oboxbox}, which enables us to clarify the size of this theoretical uncertainty in relation to the expected experimental precision. The relation of such uncertainties with theoretical power counting arguments and unitarity is clarified in passing. We offer conclusions in Sec.~\ref{sec:conc}.

\newcommand{\GeV}{\text{GeV}}
\section{Field Redefinitions vs. Equations of Motion}
\label{sec:theory}
Our approach is best illustrated initially with a simple example. Consider a scalar $h$ coupled to an external source $J$
\begin{align}\label{eq:Lh0}
    \mathcal L&=\frac12\partial_\mu h\partial^\mu h+A(h/v)J\,,& A(\hsd)&=(1+\hsd)^2+\epsilon  \hsd^2 \,,
\end{align}
where we will use $\hsd=h/v$ with $v=246~\GeV$. The parameter $\epsilon$ is taken to be small, and it modifies the coupling of two $h$-fields to the source. Let us remove such a term in the Lagrangian while keeping the same theory. This can be done by virtue of the invariance of physical observables with respect to field redefinitions. The current case field redefinition (FR) or change of scalar coordinate is
\begin{align}
\label{eq:redef}
    h=\underline h-\frac{1}{2} \epsilon \underline h^2/v\,,
\end{align}
which takes us to the new Lagrangian
\begin{align}
    \underline{\mathcal L}(\underline h)=\frac{1}{2}(\partial_\mu \underline h \partial^\mu \underline h)(1-\epsilon\underline{\hsd})^2+J\underline A(\underline h/v)\,,
\end{align}
with 
\begin{align}
    \underline A(\underline \hsd)=(1+\underline \hsd)^2-(\epsilon+\epsilon^2)\underline \hsd^3+\frac{(\epsilon^3+\epsilon^4)\underline \hsd^4}{4}\,.
\end{align}
This is a Lagrangian that, despite apparent differences, will lead to the very same amplitudes as the one in Eq.~\eqref{eq:Lh0}.

One might also remove the $\epsilon$ term in a seemingly different way, using the equation of motion (EoM) as an algebraic relation. This procedure is widely used for the definition of operator bases. It follows, $S=\int d^4x\, \mathcal L$,
\begin{align}
    \frac{\delta S}{\delta h(x)}&=-\Box h(x)+\frac{A'(\hsd)}{v} J(x)=0\,,  &J=&\frac{v}{2}\Box h-(1+\epsilon)\zeta J\,,
\end{align}
now substitution of $J$ as given above in the very term we want to remove, i.e. the $\epsilon \zeta^2 J$ term in the Lagrangian in Eq.~\eqref{eq:Lh0}, leads to
\begin{align}
    \mathcal L_{\text{EoM}}=\frac{1}{2}(\partial_\mu  h \partial^\mu  h)(1-\epsilon \hsd)+J A_{\text{EoM}}( h/v)\,,
\end{align}
with 
\begin{align}
    {A}_{\text{EoM}}(\hsd)=(1+ \hsd)^2-(\epsilon+\epsilon^2)\hsd^3\,.
\end{align}
This is yet another Lagrangian which now, however, is only guaranteed to provide the same amplitudes to order $\epsilon$. 
To understand why this is the case, we can rewrite the Lagrangian as
\begin{align}
    \mathcal L_{\text{EoM}}&=\mathcal L(h(x))+\left(-\frac{\epsilon h^2(x)}{2v}\right)\frac{\delta S}{\delta h(x)}=\mathcal L(h)+\left(-\frac{\epsilon h^2}{2v}\right)\left[-\Box h+A'(\hsd) J/v\right]\,,
\end{align}
hence using the EoM is equivalent to the first order modification in a field redefinition $\phi\to \phi+\epsilon \delta\phi$ (cf. Eq.~\eqref{eq:redef}) as highlighted in~\cite{Criado:2018sdb}. The difference between the two is therefore the second variation of the action
\begin{align}\label{eq:2nd_variation_err}
    S_{\Delta\text{TH}}=\int d^4x \,d^4y\,\frac{\epsilon^2}{2}\delta\phi(y)\frac{\delta^2 S}{\delta\phi(x)\delta\phi(x)}\delta\phi(y)\,,
\end{align}
and higher orders. This example shows the interpretation of the use of the EoM `once' to remove a given operator; one can make multiple uses of the EoM on the same operator, which is related to iterated field redefinitions. Appendix~\ref{sec:matching} shows how one can use the EoM `twice' and how the error would still take the form of Eq.~\eqref{eq:2nd_variation_err}.

Such a definition of a theory error is still a few steps away from a ready-to-use form, but the road to get there is clear: compute your observable with and without the term in the action above and take the theory error as the difference between the two. 
Let us be a bit more explicit. We define (a strictly positive)
\begin{align}
    \Delta \text{TH}=  \left|\frac{\sigma-\sigma_{\text{EoM}}}{\sigma-\sigma_{\text{SM}}}\right|\,.
    \label{eq:therr}
\end{align}
A few comments are in order. 
\begin{itemize}
\item Firstly, note there is no $\sigma_{\text{FR}}$. This is because a field redefinition truly leads to the same theory and $\sigma_{\text{FR}}=\sigma$. In practice, we will compute $\sigma_{\text{FR}}$ and compare with $\sigma$ as a cross check in Sec.~\ref{sec:oboxbox}.  
\item Secondly, the difference in the numerator stems from Eq.~\eqref{eq:2nd_variation_err} and is of order $\epsilon^2$ while the leading term in the numerator is of order $\epsilon$ so the expressions has a well-defined limit as $\epsilon\to0$ and $\sigma,\sigma_{\text{EoM}}\to \sigma_{\text{SM}}$. 
\item Lastly, this is a dimensionless and relative error, due to the balance of $\epsilon$ orders just described. It captures the goodness of the convergence in the EFT and $\Delta\text{TH}\sim 1$ signals $\text{LO}\sim\text{NLO}$ and the breakdown of the EFT expansion.
\end{itemize}

While computing this theory error now appears straightforward, it can be technically involved. So, for the example here, let us choose a simple limit in which the amplitudes are known. Take the source $J$ to be the kinetic term for the Goldstone bosons of the SM electroweak sector, 
\begin{equation}
   J=\frac{v^2}4\text{Tr}(\partial_\mu U\partial^\mu U) \,,
\end{equation}
where the unitary matrix $U$ contains the three Goldstone Bosons ($\pi^a$) of the SM and transforms as a bi-doublet under the global $SU(2)_L \times SU(2)_R$ symmetry:
\begin{align}\label{eq:goldstone}
    U(x) &\to L\, U(x)\, R^\dagger\,,&  U(x) =& \exp\left( i \frac{\sigma^a \pi^a(x)}{v} \right)\,.
\end{align}
The last equation is an explicit parametrisation, the exponential; other representations are available and just as valid. One can capture all three cases above in a Lagrangian
\begin{align}
    \mathcal L\equiv\frac12 K^2(\hsd)\partial_\mu h\partial^\mu h+ F^2(\hsd) \frac{v^2}4\text{Tr}(\partial_\mu U\partial^\mu U)-V(h)\,,
\end{align}
where the kinetic term defines a metric with line element
\begin{align}
    d\phi^2=v^2\left[K^2(\hsd) d\hsd^2+F^2(\hsd)d\Omega^2\right]\,,
\end{align}
with $d\Omega^2$ being the inner line element in $S^3$ since $U\sim S^3$. 

The high energy amplitudes for this un-gauged theory do approximate the longitudinal boson scattering of the gauged case through the equivalence theorem~\cite{Cornwall:1974km,Chanowitz:1985hj,Dobado:1990jy}, but furthermore they can be written in terms of the Riemann curvature tensor that follows from the metric as ($g_{\alpha\beta}=e_\alpha^i e_{i\beta}$) \cite{Alonso:2015fsp, Alonso:2016oah, Cohen:2021ucp,Cheung:2021yog} 
\begin{align}
    \mathcal A_{\pi_i \pi_j\to 2h}=& \frac{s}{v^2} \,e^{\alpha}_ie^{\kappa}_je^{\beta}_he^{\gamma}_h v^2R_{\alpha\beta\gamma\kappa}\,,\label{eq:RieAmp2}
\end{align}
\begin{align}
   \mathcal A_{\pi_i\pi_j\to 3h}=s\, e^{\alpha'}_h e^{\alpha}_he^{\beta}_je^{\gamma}_he^{\kappa}_\ell \nabla_{\alpha'}R_{\alpha\beta\gamma\kappa}=&-\left.\frac{1}{K(h)}\frac{dR_h}{dh}\right|_{h=0} s\, \delta_{ij}\,,\label{eq:RieAmp3}
\end{align}
where
\begin{align}
    e^{\alpha}_he^{\beta}_je^{\gamma}_he^{\kappa}_\ell R_{\alpha\beta\gamma\kappa}=&R_h \delta_{ij}\,,
\end{align}
\begin{align}
        v^2R_h&=-\frac{v^2}{K^2}\left(\frac{1}{F}\frac{d^2F}{dh^2}-\frac{1}{KF}\frac{dF}{dh}\frac{dK}{dh}\right)_{h=0}
        =-\frac{1}{K^2}\left(\frac{1}{F}\frac{d^2F}{d\hsd^2}-\frac{1}{KF}\frac{dF}{d\hsd}\frac{dK}{d\hsd}\right)_{\zeta=0}\,.
\end{align}
The form of Eqs.~\eqref{eq:RieAmp2},~\eqref{eq:RieAmp3} is that of a tensor contracted with vierbiens $T_\alpha e^{\alpha}$ and is, hence, invariant under field-coordinate transformations; there lies the usefulness of these expressions.

The matter of computing the amplitude is then reduced to identifying the Higgs-functions $K,F$ for each Lagrangian and obtaining the curvature via the expressions above. The results are collected in Tab.~\ref{tab:amplitude1}.
%
\begin{table}[!b]
    \centering
    \setlength{\arrayrulewidth}{0.5mm}
    \setlength{\tabcolsep}{18pt}
    \renewcommand{\arraystretch}{1.6}
    \begin{tabular}{m{.5cm}  m{2cm} m{3.5cm} m{3.5cm}  } 
        \hline
        & $\mathcal L$ & $\underline{\mathcal L}$ & $\mathcal L_{\text{EoM}}$ \\
        \hline
        $K^2(\hsd)$& $1$ & $1-2\epsilon \hsd+\epsilon^2 \hsd^2$ & $1-2\epsilon \hsd$ \\
        $F^2(\hsd)$ & $ (1+\hsd)^2+\epsilon \hsd^2 $ & $(1+\hsd)^2-(\epsilon+\epsilon^2)\hsd^3+ O(\hsd^4)$ & $(1+\hsd)^2 - (\epsilon+\epsilon^2)\hsd^3+ O(\hsd^4)$\\
        $v^2R_h$ & $-\epsilon$ &$-\epsilon$ &$-\epsilon$ \\[0.2cm]
        $\displaystyle{\frac{v^3}{K}\frac{dR_h}{dh}}$ & $4\epsilon$ & $4\epsilon$ & $4\epsilon-\epsilon^2$ \\[0.3cm]
        \hline
    \end{tabular} 
    \caption{Metric and curvature elements for the original, field-redefined and EoM-transformed Lagrangians. The curvature terms give the high-energy leading terms in amplitudes and signal physical differences between theories.}
    \label{tab:amplitude1}
\end{table}
%
Coincidentally, the $2\to 2$ scattering terms happen to have only linear $\epsilon$ corrections, which means all three Lagrangians give the same prediction. The $2\to3$ case, however, does display both the equivalence of the original and field-redefined case and the fact that EoM lead to a different prediction at order $\epsilon^2$.

For the $\pi\pi\to 3h$ case, we can further identify the ratio between LO prediction and NLO theory error as $\epsilon^2/4\epsilon=\epsilon/4$. If a hypothetical experiment would only set limits in the range $\epsilon \sim 4$, e.g. when statistical control is not (yet) good enough, no reliable information can be gained from this measurement when it is interpreted in a truncated EFT approach. Our example, therefore, clearly highlights the tension between data quality and EFT expansion parametrised by the theory error of Eq.~\eqref{eq:therr}.

Generalising this example requires examining two issues: firstly, the general case will not be the high-energy limit of a scalar theory or have simple geometry-based formulae to connect the Lagrangian with observables. And secondly, there is the issue of determining our expansion parameter $\epsilon$, either via theoretical input or directly via measurements as indicated above.

The first point is no obstacle; we simply have to specify our theory in full and the operator whose removal will lead to the $\mathcal{O}(\epsilon^2)$ error. The full theory we choose as HEFT, which requires gauging $SU(2)_L\times  U(1)_Y$, so that
\begin{equation}
    D_\mu U = \partial_\mu U + i \frac{g}{2} W_\mu^I \sigma_I U - i g' U B_\mu T_Y\,,
\end{equation}
where $T_Y$ is the generator of $U(1)_Y$ in scalar space, explicitly $T_Y=\sigma_3/2$. 

The part of the HEFT Lagrangian (considering an expansion in chiral dimensions) that has been tested experimentally is contained in
\begin{equation}
\label{eq:heft}
    \lag_{\text{HEFT}} = \lag_{\text{bosonic}} + \lag_{\text{fermion}} + \lag_{\text{Yukawa}} + \lag^{(4)}_{\text{HEFT}}\,,
\end{equation}
were $\lag^{(4)}_{\text{HEFT}}$ refers to the chiral dimension four (next-to-leading order) Lagrangian. The leading order parts of the HEFT Lagrangian are given by
\begin{align}
    \lag_{\text{bosonic}} &= \frac{1}{2} \partial_\mu h \partial^\mu h + \frac{v^2}{4} \mathcal{F}_C(h) \, \text{Tr}\big[(D_\mu U)^\dagger D^\mu U\big] - V(h) \nonumber \\
    &\quad - \frac{1}{4} W_{\mu\nu}^a W^{a \mu\nu} - \frac{1}{4} B_{\mu\nu} B^{\mu\nu}\,, \\
    \lag_{\text{Yukawa}} &= - 
    \frac{v}{\sqrt{2}} \sum_f \big( \bar{\psi}_L U \, \mathcal{Y}_f(h) \, \psi_R + \text{h.c.} \big)\,, \\
    \lag_{\text{fermion}} &= \sum_\psi \bar{\psi} i \slashed{D} \psi\,.
\end{align}
The function $\mathcal{F}_C(h)$ describes possible modifications of the Higgs boson couplings to gauge bosons and can be expanded as a polynomial in $h$:
\begin{equation}
    \mathcal{F}_C(h) = 1 + 2 a \frac{h}{v} + b \frac{h^2}{v^2} + O\left(\frac{h^3}{v^3}\right)\,,
\end{equation}
while the Higgs-dependent Yukawa couplings take the form
\begin{equation}
    \mathcal{Y}_f(h) = y_f \left(1 + c_f \frac{h}{v} + O(h^2/v^2) \right)\,,
\end{equation}
where the sum runs over fermion species. The scalar potential is also kept general,
\begin{equation}
    V(h) = \frac{1}{2} m_h^2 h^2 + \kappa_3 \frac{m_h^2}{2 v} h^3 + \kappa_4 \frac{m_h^2}{8 v^2} h^4 + O(h^5)\,.
\end{equation}

To be specific, when we say that this Lagrangian has been tested, we mean that the coefficients of the $h-$functions have been measured up to linear order (except for the lightest generations' Yukawa couplings), while for the potential, we have an upper bound on the cubic term. 
The Lagrangian is grouped into such operators because Lorentz and gauge invariance restrict its possible terms to this form. In particular, arbitrary functions of $h$ are allowed, since $h$ does not transform under the gauge group.

We now come back to the second point, the question of determining the theory's expansion parameter. Finding operators invariant under Lorentz and gauge symmetry has a known solution expressed through Hilbert series~\cite{Henning:2015daa, Alonso:2024usj,Lehman:2015via, Lehman:2015coa, Henning:2015alf, Henning:2017fpj, Graf:2020yxt, Graf:2022rco, Sun:2022aag,Hanany:2010vu, Banerjee:2020bym}. An issue that is less straightforward and more subject to discussion is how to `order' these operators, i.e. finding our $\epsilon$. A limit with unambiguous expansion is SMEFT, in which the assembly of $U$ and $h$ into a doublet $H$ provides all operators beyond the SM with a mass scale $\Lambda$, by assumption greater than the electroweak vacuum expectation value $v\simeq 246~\GeV$. One therefore has an expansion parameter $\epsilon\sim\Lambda^{-2}$ and an error captured by a dimension eight operator. The connection with analyses of SMEFT such as~\cite{Hays_2019,Hays_2020} can be made explicit matching to the operators that would yield the curvature of Tab.~\ref{tab:amplitude1}. The curvature-operator relation was worked out in \cite{Alonso:2021rac} and the dimension eight operator in the case at hand would be a combination of $H^\dagger H [\partial (H^\dagger H)]^2$ and $(H^\dagger H)^2D_\mu H^\dagger D^\mu H$.

In HEFT, no such simple identification of $\epsilon$ is possible; although the SM limit is straightforward to take, this is not achieved by sending mass-dimension $\ge 4 $ operators to $0$. This can be understood as HEFT can interpolate between technicolour-like theories, where the cut-off is of order $4\pi v$, and SMEFT-like theories. To fix this uncertainty, one can choose a set of operators that are taken as the leading order (LO) Lagrangian and determine the rest of the series by the loop order at which they are generated. This line of thought is to be found in~\cite{Buchalla:2013eza}. One can also use Naive Dimensional Analysis (NDA) as a guide to normalise operators which have a scale $\Lambda$ and $4\pi$ factors and coefficients $c_i$ such that the perturbative loop expansion breaks down at $c_i\simeq 1$, as done in~\cite{Gavela:2016bzc,Brivio:2025yrr,Brivio:2025sib}. 

Our approach, detailed above, offers another way to define the limit of validity, in which the error becomes comparable with the prediction. This is, in spirit, the same procedure as changing the renormalisation scale to assess theory uncertainties. Following this, we discuss how one can quantify such a truncation error with an explicit example in the remainder of this work.

\section{A $\cal{O}_{\Box\Box}$ Case Study}
\label{sec:oboxbox}
We now turn to a concrete example in the Higgs sector that upgrades the instructive case study of the previous section to a distinctive HEFT operator whose impact at colliders can be examined. The operator we focus on is
\begin{equation}
\label{eq:opdef}
    \obb = \square h \, \square h\,,
\end{equation}
which is known to prove characteristic in HEFT vs. SMEFT comparisons~\cite{Brivio:2014pfa,Anisha:2024xxc}. Such non-canonical two-point-function contributions arise naturally in HEFT, e.g. through loop corrections (see~\cite{Herrero:2022krh}) and they modify the Higgs boson's propagation, with measurable effects in scattering experiments. $\obb$ also lends itself to field and EoM re-definitions as detailed in the previous section (including the application of EoMs more than once). It is therefore a well-motivated test case to understand associated theoretical uncertainties when contrasted with existing and future Higgs property measurements. Independent of the theoretical uncertainty motivated in the previous section, any EFT interaction (including the momentum-dependent one we consider in the following) is subject to unitarity bounds. We will return to this for $\obb$ in relation to the processes we consider further below.

We set the stage by splitting the Lagrangian into two pieces, which will be particularly useful for the study of the considered interaction
\begin{align}
    \mathcal L&=\mathcal L_{2\text{pt}}^h+\mathcal{L}_{\text{int}}^h\,,
    \label{eq:2ptInt}
\end{align}
where $\obb$ will be included in $\mathcal L_{2\text{pt}}^h$.
 
\subsection{Lagrangians and Amplitudes}
\subsubsection{The `Standard' HEFT Framework}
\label{sec:vanilla}
We now turn our attention to the specific role of the operator $\obb$ within a `vanilla' HEFT framework as a representative example to discuss theoretical uncertainties when confronted with data. Following the approach of Ref.~\cite{Herrero:2022krh}, $\obb$ appears in the chiral dimension four Lagrangian as
\begin{equation}
    \label{eqn:aboxbox}
    \lag_{2\text{pt}}^h \supset -\frac{\abb}{v^2} \obb\,.
\end{equation}
This aligns our sign convention. In Eq.~\eqref{eq:heft}, it is contained in the last term, $\lag_{\text{HEFT}}^{(4)}$.
The Lagrangian term that produces the two-point function in this theory then reads
\begin{equation}
    \lag_{2\text{pt}}^h = \frac{1}{2} \left( \partial_\mu h \partial^\mu h - m_h^2 h^2 - \frac{2 \abb}{v^2} \square h \, \square h \right)\,.
\end{equation}
The Higgs propagator can then be calculated from the EoM of the free Higgs field, which is solved perturbatively by the associated Green's function ($G_h$) in momentum space,
\begin{equation}
    iG_h^{-1} (p^2) = p^2 - m_h^2 - \frac{2 \abb}{v^2} p^4\,.
    \label{eqn:propvanilla}
\end{equation}
This modification changes the location of the pole Higgs mass\footnote{Our discussion is limited to the leading order. It is nonetheless interesting to point out that the location of unstable particles like the Higgs boson resides in the second Riemann sheet of the $S$-matrix: The complex nature of the pole is critical beyond leading order to maintaining gauge-invariance~\cite{Passarino:2010qk,Gambino:1999ai} of the Higgs resonance as a definition of a signal in the SM~\cite{Goria:2011wa} and, hence, all its extensions. On the contrary, in HEFT, with the physical Higgs arising as an iso-singlet state, no such reservations apply.} as can be seen by rewriting
\begin{align}
  i  G^{-1}_h(p^2)&=\left(1-2\frac{\abb}{v^2}\left[p^2+m_{\text{ph}}^2\right]\right)\left(p^2-m_{\text{ph}}^2\right)\,,
\end{align}
which is the same as Eq.~\eqref{eqn:propvanilla} to order ${O}(\abb^3)$, and where  
\begin{align}
\label{eq:massrel}
m_{\text{ph}}^2&=m_h^2\left(1+2\abb \frac{m_h^2}{v^2}+8 \abb^2\frac{m_h^4}{v^4}\right)\,, & m_h^2&=m_{\text{ph}}^2\left(1-2\abb\frac{m_{\text{ph}}^2}{v^2}\right)\,.
\end{align}
Additionally, the pole-residue is now non-canonical, so we renormalise our field as
\begin{align}
    h&=Z^{1/2}h_R\,, & Z&=1+4\abb\frac{m_{\text{ph}}^2}{v^2}+16\abb^2\frac{m_{\text{ph}}^4}{v^4}\,.
\end{align}
In this basis, we have the renormalised Higgs propagator
\begin{align}
    G_R(p^2)=\frac{i}{p^2-m_{\text{ph}}^2}\left(1+2\frac{\abb}{v^2} (p^2+m_{\text{ph}}^2)+4\frac{\abb^2}{v^4} (p^2+m_{\text{ph}}^2)^2\right)\,.
\end{align}
The interaction Lagrangian is then modified as
\begin{multline}
    \lag_{\text{int}}^{h_R}=\lag_{\text{int}}^{h}(Z^{1/2}h_R)=\lag_{\text{int}}^{h}(0)+\frac{\partial \lag_{\text{int}}^{h}(0)}{\partial h}\left(1+2\abb\frac{m_{\text{ph}}^2}{v^2}+6\abb^2\frac{m_{\text{ph}}^4}{v^4}\right)h_R+ O(h_R^2)\,,
\end{multline}
which implies a rescaling of all Higgs couplings with respect to the SM case. To be more explicit, and in anticipation of the processes discussed further below, we can select the Higgs-mediated fermion-fermion scattering process $f (p_f) \bar f (p_{\bar{f}}) \rightarrow h \rightarrow f^\prime(p_{f^\prime}) \bar{f^\prime} (p_{\bar{f^\prime}} )$ we have
\begin{multline}
\parbox{3cm}{
    \begin{tikzpicture}
        \begin{feynman}
            \vertex (a) at (0, 0);
            \vertex (b) at (1, 0);
            \vertex (t1) at (-0.7, 0.7) {\(\large f\)};;
            \vertex (t2) at (-0.7, -0.7) {\(\large \bar{f}\)};;
            \vertex (b1) at (1.7, 0.7) {\(\large f^\prime\)};
            \vertex (b2) at (1.7, -0.7) {\(\large \bar{f^\prime}\)};        
            \diagram* {
                (a) -- [scalar, edge label = \(\large h\)] (b),
                (t1) -- [fermion] (a) -- [fermion] (t2),
                (b1) -- [anti fermion] (b) -- [anti fermion] (b2),
            };
        \end{feynman} 
    \end{tikzpicture}
}
\sim\mathcal{M}^{\text{vanilla}} \\= \frac{-i y_f y_f'/2}{s-m_\text{ph}^2}\left(1+2\frac{\abb}{v^2} (s+m_{\text{ph}}^2)+4\frac{\abb^2}{v^4} (s+m_{\text{ph}}^2)^2\right)\mathcal A_{ff'}\,,\label{eq:vanfactors}
\end{multline}
where
\begin{equation}
    \mathcal A_{ff'}=\left[ \bar{v}(p_{\bar{f}}) P_R u(p_f) \right]
\left[ \bar{u}(p_{f^\prime}) P_L v(p_{\bar{f^\prime}}) \right]
\end{equation}
contains all the spinors, $P_{L,R} = (1 \mp \gamma^5)/{2}$ are the left and right-chiral projection operators, $y_{f,f^\prime}$ are the corresponding Yukawa couplings of the Higgs to $f,f^\prime$, and $s=(p_f + p_{\bar{f}})^2=(p_{f^\prime} + p_{\bar{f^\prime}})^2$ the squared centre-of-mass energy. 
Note in particular that for an on-shell Higgs kinematics
\begin{align}
     \left.\mathcal{M}^{\text{vanilla}}\right|_{\text{on-shell}} 
     = \left( 1 + \frac{4 \abb}{v^2} m_{\text{ph}}^2 + \frac{16\abb^2}{v^4} m_{\text{ph}}^4 \right)  \times \left.\mathcal{M}^{\text{SM}}\right|_{\text{on-shell}}\,.
\end{align}

\subsubsection{The Field-Redefined HEFT Lagrangian}
We now perform a perturbative field redefinition of the Higgs field to remove the higher-derivative operator $\obb$ from the chiral dimension four Lagrangian. The original Higgs `coordinate' reads, in terms of the new one,
\begin{equation}
    h = h^{\prime} - \frac{\abb}{v^2} \Box h^{\prime}\,.
\end{equation}
This cancels the $\obb$ operator, yet it also generates other interactions as
\begin{equation} \label{eq:Leff_fr}
    \lag^{h'}_{2\text{pt}} = \frac{1}{2} \partial_\mu h^{\prime} \partial^\mu h^{\prime} - \frac{1}{2} m_h^2 h^{\prime 2} + m_h^2 \frac{\abb}{v^2} h^{\prime} \Box h^{\prime} - \frac{1}{2} m_h^2 \frac{\abb^2}{v^4} \Box h^{\prime} \Box h^{\prime}
    + \frac{3}{2} \frac{\abb^2}{v^4} \Box h^{\prime} \Box^2 h^{\prime}\,.
\end{equation}
The resulting inverse Green function is
\begin{equation}
    i\,G_h^{\prime-1}(p^2) = p^2 - m_h^2 - \frac{2 \abb}{v^2} m_h^2 p^2 - \frac{\abb^2}{v^4} m_h^2 p^4 - \frac{3 \abb^2}{v^4} p^6\,.
\end{equation}
The procedure to obtain a canonical field is the same as above; one rewrites
\begin{align}
    i\,G_h^{\prime-1}=\left(1-2\frac{\abb}{v^2} m_{\text{ph}}^2-\frac{\abb^2}{v^4}p^2(3p^2+4m_{\text{ph}}^2)\right)(p^2-m_{\text{ph}}^2)\,,
\end{align}
where the mass relation is the same as in Eq.~\eqref{eq:massrel}. The field renormalisation is
\begin{align}
    h'&=(Z')^{1/2}h'_R\,,&Z'&=1+2\frac{\abb}{v^2} m_{\text{ph}}^2+11 \frac{\abb^2}{v^4} m_{\text{ph}}^4\,,
\end{align}
and the renormalised propagator takes the form
\begin{align}
    G_R'(p^2)=\frac{i}{p^2-m_{\text{ph}}^2}\left(1+\frac{\abb^2}{v^4}(3 p^4+4p^2 m_{\text{ph}}^2-7m_{\text{ph}}^4)\right)\,.
\end{align}
The couplings of the renormalised field are
\begin{align}
    \lag_\text{int}^{h'_R}=&\lag_\text{int}^h\left[\left(1-\frac{\abb}{v^2}\Box)(Z')^{1/2}h_R'\right) \right]\nonumber\\=&
    \lag_{\text{int}}^{h}(0)
    +\frac{\partial \lag_{\text{int}}^{h}(0)}{\partial h}\left(1+
    \abb\frac{m_{\text{ph}}^2-\Box}{v^2}+\abb^2m^{2}_{\text{ph}}\frac{5m_{\text{ph}}^2-\Box}{v^4}\right)h_R'+ O(h_R^{\prime 2})\,.
\end{align}
In momentum representation, the $\Box$ operator will return $-p_h^2$, giving again a rescaled set of couplings for the Higgs but now they depend on the Higgs boson's kinematics.

The amplitude for the same process of Eq.~\eqref{eq:vanfactors} now reads
\begin{equation}
    \frac{\mathcal{M}^{\text{FR}}}{\mathcal{M}^{\text{SM}}} (f \bar{f} \to h \to f^\prime \bar{f^\prime}) = 
    1+2\frac{\abb}{v^2} (s+m_{\text{ph}}^2)+4\frac{\abb^2}{v^4} (s+m_{\text{ph}}^2)^2
    \,,
    \label{eqn:frfactors}
\end{equation}
that is, exactly the same as the original one, thus providing an independent check of our derivation.

\subsubsection{Equation of Motion Substitution}
Let us write the EoM as
\begin{align}
    \Box h=-m_h^2h -2 \frac{\abb}{v^2} \Box^2h+\frac{\partial\lag_{\text{int}}^h}{\partial h}\,,
\end{align}
we shall use it to substitute one of the $\Box h$ terms in $\obb$. The new two-point Lagrangian with this use of the EoM reads
\begin{equation} \label{eq:Leff_eom}
    \mathcal{L}_{2\text{pt}}^{\text{EoM}} = \frac{1}{2} \partial_\mu h \partial^\mu h - \frac{1}{2} m_h^2 h^2 + \frac{\abb}{v^2} m_h^2 h \boxh +\frac{2 \abb^2}{v^4}\Box h \Box^2 h\,,
\end{equation}
where we note that the difference of Eq.~\eqref{eq:Leff_fr} and Eq.~\eqref{eq:Leff_eom} is
\begin{equation}
    \frac12\frac{\abb}{v^2} \Box h (-m_h^2-\Box)\frac{\abb}{v^2}\Box h\,.
\end{equation}
This is the form of Eq.~\eqref{eq:2nd_variation_err} when restricted to two Higgs terms, as is the case here. The inverse propagator in the EoM theory is
\begin{equation}
    i\,(G^{\text{EoM}}_h)^{-1}(p^2) = p^2 - m_h^2 - \frac{2 \abb}{v^2} p^2 m_h^2 - \frac{4 \abb^2}{v^4}p^6\,,
\end{equation}
which in factorised form equals
\begin{align}
     i \,(G^{\text{EoM}}_h)^{-1}=\left(1-2\frac{\abb}{v^2}m_{\text{ph}}^2-4\frac{\abb^2p^2}{v^4}\left(p^2+m_{\text{ph}}^2\right)\right)\left(p^2-m_{\text{ph}}^2\right)\,,
\end{align}
and requires renormalisation
\begin{align}
     h&=Z_{\text{EoM}}^{1/2}h_R\,,&Z_{\text{EoM}}=&1+2\frac{\abb}{v^2} m_{\text{ph}}^2+16 \frac{\abb^2}{v^4} m_{\text{ph}}^4\,.
\end{align}
One then has all the pieces to put together perturbative computations, and the propagator~is
\begin{align}
    G_R^{''}=\frac{i}{p^2-m_{\text{ph}}^2}\left(1+\frac{4 \abb^2}{v^4}\left(p^4+p^2m_{\text{ph}}^2-2m_{\text{ph}}^4\right)\right)\,,
\end{align}
while the couplings of the Higgs boson are now
\begin{align}
    \lag_{\text{int}}^{\text{EoM},h_R}=&\lag_{\text{int}}^{h}(Z_{\text{EoM}}^{1/2}h_R)-\abb (\Box Z_{\text{EoM}}^{1/2}h_R )\frac{\partial\lag_{\text{int}}^h(Z_{\text{EoM}}^{1/2}h_R)}{\partial h}\nonumber\\
    =&\lag_{\text{int}}^{h}(0)+\frac{\partial\lag_{\text{int}}^h(0)}{\partial h}\left(1+\abb\frac{m_{\text{ph}}^2-\Box}{v^2}+\abb^2m_{\text{ph}}\frac{11m_{\text{ph}}^2-2\Box}{2v^4}\right)h_R+ O(h_R^2)\,,
\end{align}
with a kinematic-dependent rescaling of couplings as in the field-redefinition case. Turning finally to our amplitude of choice, we find
\begin{equation}
    \frac{\mathcal{M}^{\text{EoM}}}{\mathcal{M}^{\text{SM}}} (f \bar{f} \to h \to f^\prime \bar{f^\prime}) = 
    \left(1+2\frac{\abb}{v^2} (s+m_{\text{ph}}^2)+\frac{4\abb^2}{v^4} (s+m_{\text{ph}}^2)^2\right)+\frac{\abb^2}{v^4}s^2\,.
    \label{eqn:EoMfactor}
\end{equation}
in the EoM theory. For on-shell Higgs kinematics (relevant for Higgs particle production and decay), this becomes
\begin{equation}
    \left.\frac{\mathcal{M}^{\text{EoM.}}}{\mathcal{M}^{\text{SM}}} \right|_{\text{on-shell}} = 
    1 + \frac{4 \abb}{v^2} m_{\text{ph}}^2 + \frac{17 \abb^2}{v^4} m_{\text{ph}}^4 \,.
\end{equation}
The EoM case has the same linear dependence but a different $\abb^2$ term as expected. Incidentally, the on-shell difference is small, $16$ vs $17$, whereas off-shell, we have an additional additive factor proportional to $s^2$. In processes that probe precisely the momentum-dependence of interactions, one of the key targets of effective field theory measurements, the reliance on EoMs to construct minimal operator sets can introduce a serious flaw in obtaining robust constraints. This crucially depends on the measurement's experimental precision, and we will clarify this interplay in the next section.

\subsection{Measurements, Constraints, Uncertainties}
We are now equipped to compare the form of the physical amplitude in the three cases: the vanilla HEFT, the field-redefined HEFT, and the EoM-substituted scenario. For processes that are sensitive to the on-shell Higgs propagator and vertex corrections at the 10\% level or below, the resulting constraints are expected to be similar across all formulations.

We approach our application to phenomenology strategically. The discussion of the previous section demonstrated that uncertainties can impact on- and off-shell regions differently. This is particularly relevant when the processes on which we rely to set constraints on a particular HEFT parameter are predominantly probed in, e.g., off-shell Higgs production. Hence, we put the discussion of the previous section into the context of two concrete examples. Firstly, we consider a $\abb$ fit to combined on-shell Higgs data. These processes have been measured to very good precision already, and we can expect the high luminosity (HL) LHC data set to achieve considerable additional sensitivity~\cite{Cepeda:2019klc}, especially when differential information can be further exploited~\cite{Belvedere:2024wzg}. Secondly, we consider the production of 4 top quarks that shows great sensitivity its electroweak production contribution~\cite{Alvarez:2016nrz,Frederix:2017wme,Alvarez:2019uxp,Blekman:2022jag} which can be exploited to set constraints on non-standard Higgs momentum dependencies~\cite{Anisha:2023xmh}. Albeit we can expect to increase the sensitivity to this process considerably beyond early extrapolations~\cite{Belvedere:2024wzg}, the uncertainty of the measurement will remain statistically challenging. This example, therefore, provides an avenue to analyse the above error in an environment where the data uncertainty is large and the sensitivity arises from off-shell Higgs kinematics.

\subsubsection{Higgs Signal Strength Data Fit}
\begin{figure}[!t]
    \centering
    \includegraphics[width=0.52\linewidth]{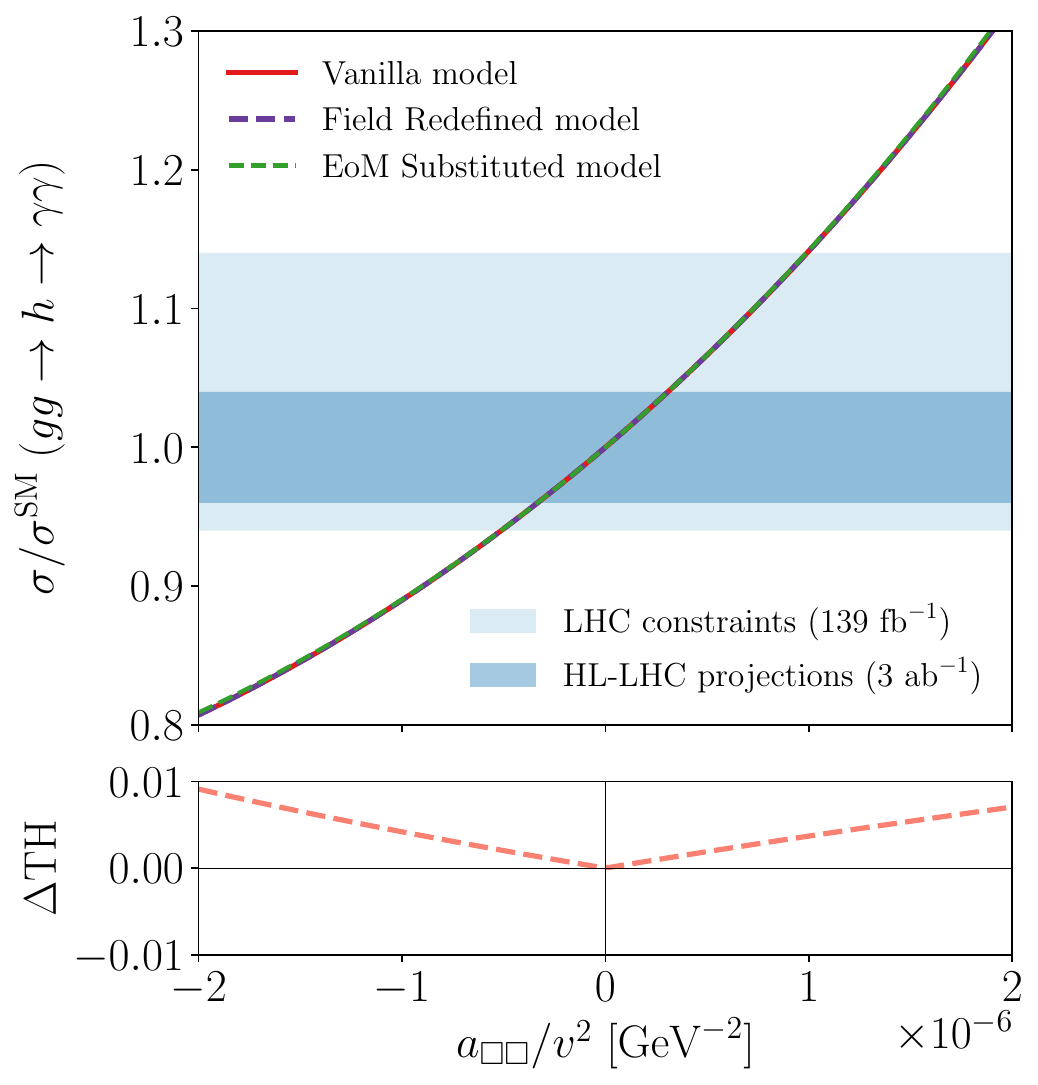}
    \caption{Ratio of the BSM-modified to SM cross section for $gg \to h \to \gamma\gamma$ at $\sqrt{s} = 13~\text{TeV}$, as a function of $a_{\square\square}/v^2$. Theoretical variations corresponding to the vanilla HEFT model, field redefinition, and EoM reduction are shown. Shaded bands indicate current (LHC, integrated luminosity of $139~\text{fb}^{-1}$~\cite{ATLAS:2022vkf}) and projected (HL-LHC, integrated luminosity of $3~\text{ab}^{-1}$~\cite{Cepeda:2019klc}) experimental sensitivities at 95\% CL. The lower panel shows the theory error calculated using Eq.~(\ref{eq:therr}).\label{fig:ggHgammagamma} }
\end{figure}

\ce{First,} we consider signal strength and cross-section measurements of the Higgs boson. As a first step, we examine the constraints on the operator $\mathcal{O}_{\square \square}$ from the process $g g \to h \to \gamma \gamma$. This channel is straightforward to evaluate using \feynarts, \formcalc, and \packagex, applied to the different HEFT Lagrangians discussed earlier. The results are shown in Fig.~\ref{fig:ggHgammagamma}, together with the $95\%$ confidence level constraints from Higgs signal strength measurements~\cite{ATLAS:2022vkf,Cepeda:2019klc}.

As expected, the relatively tight experimental bounds on this process translate into strong (data-informed) constraints on $\abb$. While the quadratic contributions in $\abb$ lead to minor differences between the EoM-substituted and field-redefined (or vanilla) cases, the linear approximation already captures the dominant effect. All three formulations, therefore, yield comparable constraints at the same order of magnitude, indicating that the field redefinition and EoM substitution are equivalent for this observable at leading order. The theoretical uncertainty, shown in the lower panel of Fig.~\ref{fig:ggHgammagamma}, exhibits an almost linear behaviour in the vicinity of the origin, as anticipated in our earlier discussion, but not trivially so, given the processing that the theory has undergone to arrive at these predictions. The error is therefore well controlled within the experimentally allowed region, reinforcing the robustness of the linear approximation for this process.

\begin{figure}[!t]
    \centering
    \includegraphics[width=0.48\linewidth]{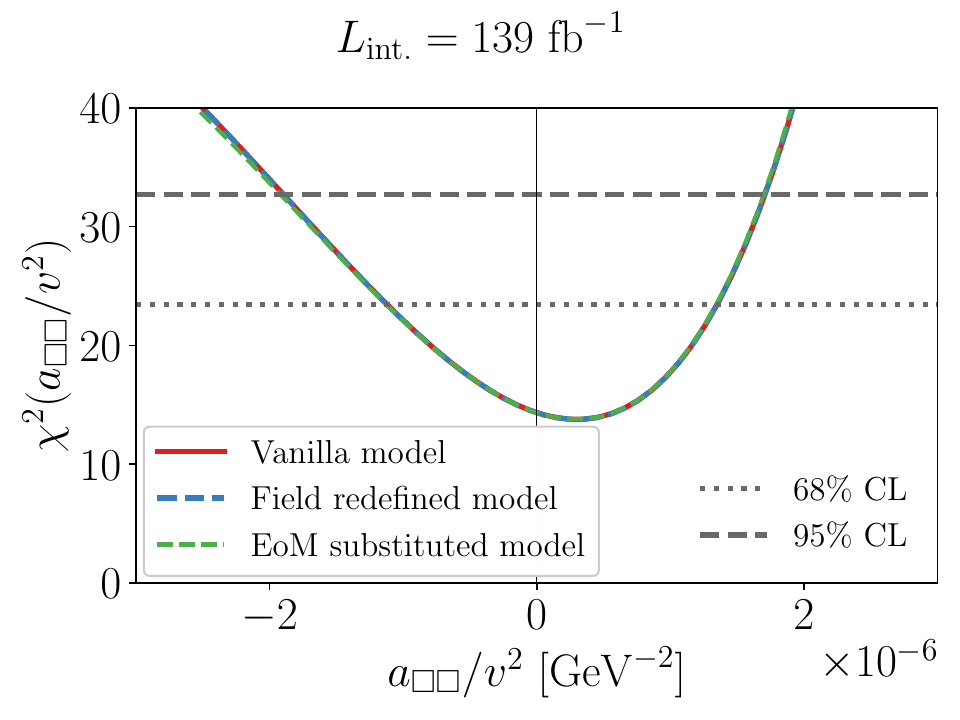}
    \includegraphics[width=0.48\linewidth]{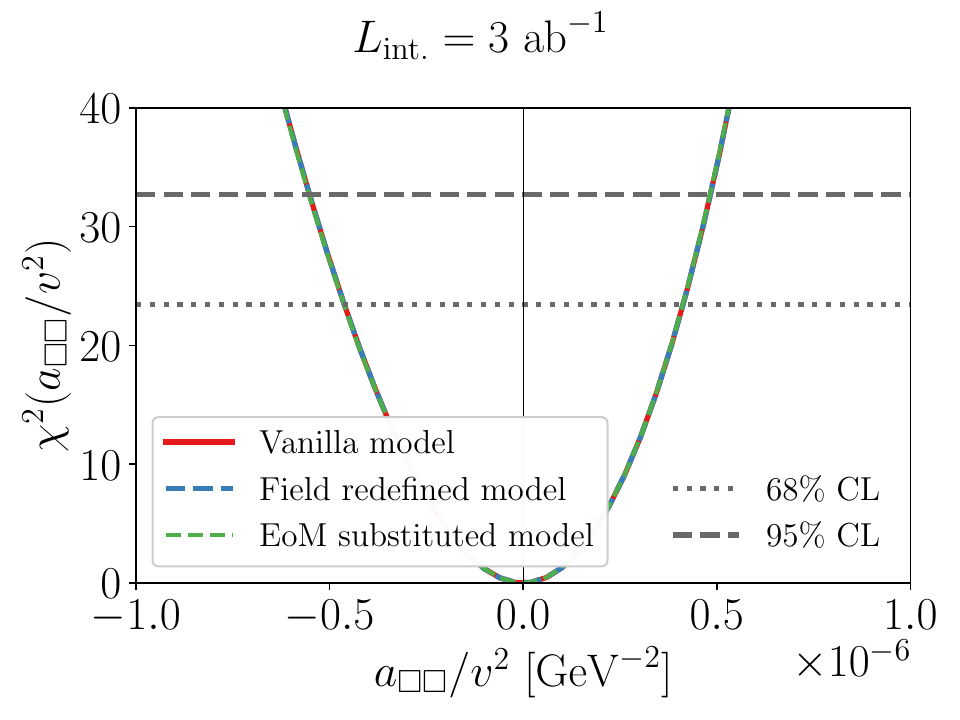}
    \caption{The $\chi^2$ fit to $\abb$ from Higgs signal strength data~\cite{ATLAS:2022vkf} (left), and the resulting extrapolation to the High Luminosity (HL-LHC) frontier (right, assuming a SM outcome) for the different HEFT Lagrangians described in the text. The $68\%$ and $95\%$ constraints are shown by the dotted and dashed black lines on the plots, respectively. \label{fig:higgsdata}  }
\end{figure}

Going further, we extract constraints on $\abb/v^2$ by performing a global fit across all relevant Higgs production and decay channels in single Higgs processes at the LHC, and at a future HL-LHC. To this end, we construct a $\chi^2$ statistic incorporating the experimental correlation matrix from Ref.~\cite{ATLAS:2022vkf}. The resulting bounds on $\abb$ are shown in Fig.~\ref{fig:higgsdata}, further illustrating that all three approaches lead to consistent results when computing physical observables at leading order.

\subsubsection{Off-shell Effects in Four-Top Production}
\label{sec:4top}
We turn to our second example, the production of four top quarks at hadron colliders,
\begin{equation}
    pp \to t\bar{t}t\bar{t}\,.
\end{equation}
This is a rare but highly informative process. Although the SM predicts a small cross section of about 13~fb at $\sqrt{s} = 13~\text{TeV}$~\cite{ATLAS:2023ajo,CMS:2023ftu}, this channel is sensitive to a wide range of potential new physics effects and has attracted increasing interest in recent years.
\refstepcounter{figure}
\begin{figure}[!t]
\begin{minipage}{1\textwidth}
\begin{small}
\begin{wrapfigure}[6]{r}{0.48\textwidth}
  \centering
  \vspace{-2.84cm}
  \includegraphics[width=\linewidth]{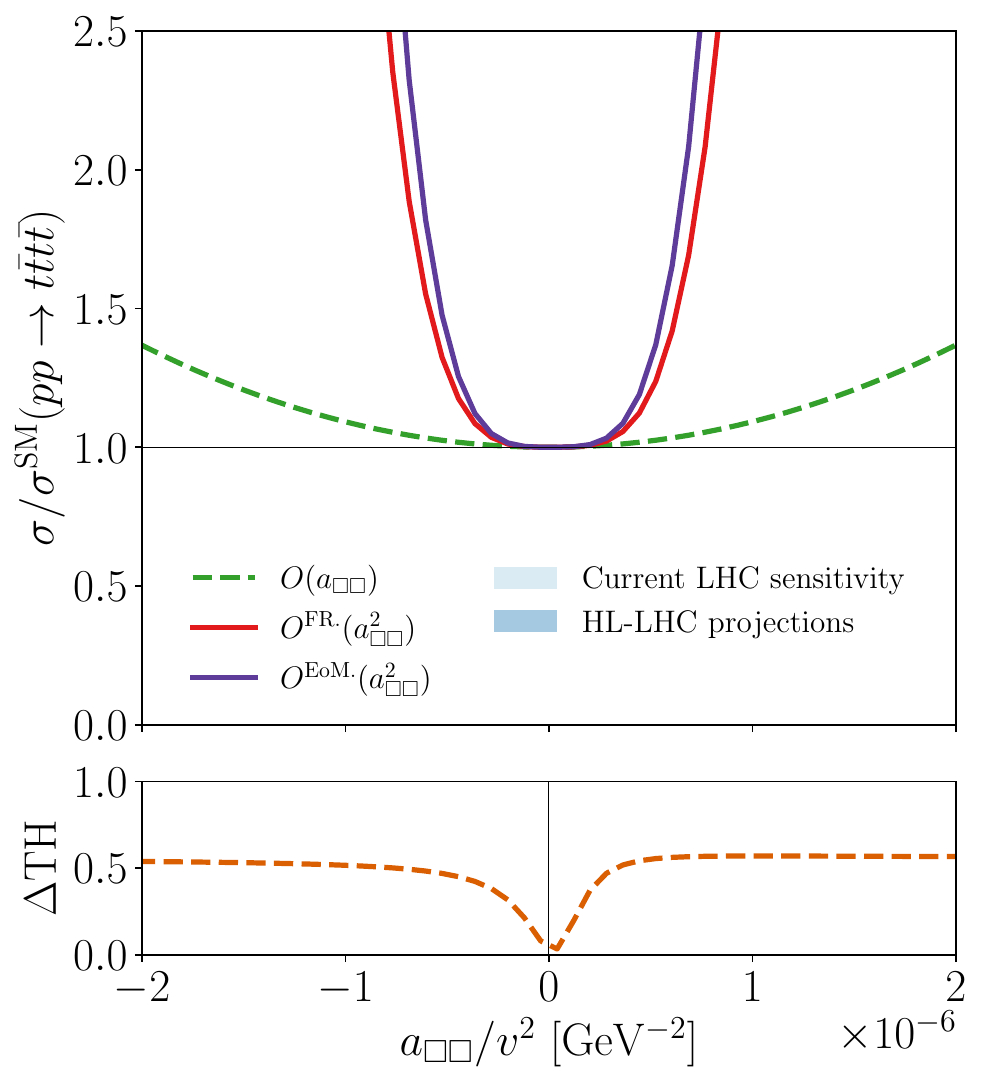}
\end{wrapfigure}
\begin{minipage}{0.45\textwidth}
 \centering \includegraphics[width=1\linewidth]{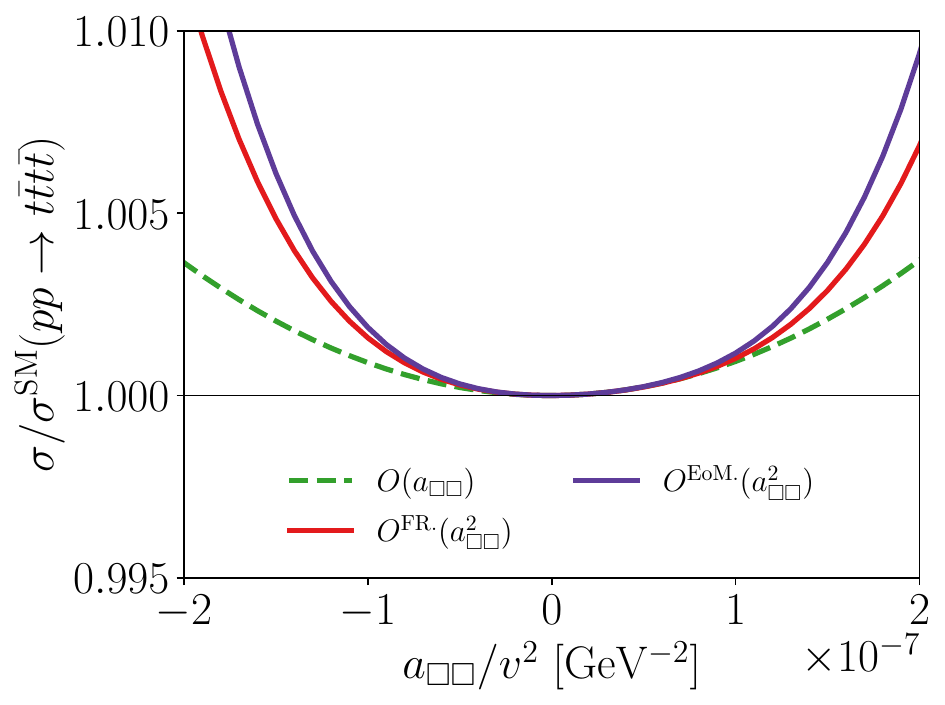}
\end{minipage}
\vskip 0.4cm
$\text{\bf{Figure~3}}$. Cross section dependence of $p p \to t \bar t t \bar t$ at $\sqrt{s} = 13$~TeV with $\abb$ for the vanilla and field redefined case for linear and quadratic truncations in the amplitudes. The light and dark bands in blue present the $95\%$ confidence limits in the current (at an integrated luminosity of $139~\text{fb}^{-1}$) and HL-LHC projected (at $3~\text{ab}^{-1}$) sensitivities to the 4-top signal strength. The lower panel on the right plot illustrates the theory errors associated with the different parameterisations described previously. The plot on the left provides a zoomed-in version of the cross section scaling on the right, in the regime where the theory error scales linearly. \label{fig:4topxs}

\end{small}
\end{minipage}
\end{figure}
Among the subleading contributions to this process, Higgs-mediated diagrams play a distinctive role. While QCD dominates the total cross section, also electroweak effects enter at the amplitude level, and are significant. Diagrams featuring a virtual Higgs exchanged between top-quark lines offer direct sensitivity to the structure of the Higgs sector~\cite{Englert:2019zmt,Banelli:2020iau}, particularly at high energies where such exchanges are far from resonance and sensitive to the detailed form of the propagator. The non-resonant nature of this process makes it well-suited to probing deviations from the SM, especially those that alter the high-energy behaviour of the Higgs exchange. Within the HEFT framework, this makes four-top production an ideal probe of higher-derivative operators such as $\obb = \obxbx$, which modify the Higgs boson's kinetic structure and affect its propagation in a momentum-dependent way. The effects of kinematic enhancement can be further amplified by the multiple operator insertions. In fact, it turns out that the linearised amplitude is relatively suppressed relative to its quadratic dependence $\sim\abb^2$. In parallel to quantifying the HEFT error, we will also contextualise its size for quadratic insertions, i.e. the amplitude is expanded up to terms $\sim \abb^2$ and analysed according to the discussion of the previous section.

To quantify the impact of $\obb$ on this channel, we evaluated both linear and quadratic contributions to the amplitude. Cross sections were generated using \texttt{MadGraph5}\_\texttt{aMC@NLO}~\cite{Alwall:2011uj}, with the modified propagator structures from Eqs.~\eqref{eq:vanfactors} and~\eqref{eqn:frfactors} implemented directly via manual edits to the \texttt{Helas} routines~\cite{Murayama:1992gi}. This ensured a consistent treatment of interference and squared terms originating from Higgs-mediated contributions. The resulting changes in the total cross section are shown in Fig.~\ref{fig:4topxs}, where a clear distinction between the linear and quadratic effects of $\obb$ can be seen. 
Figure~\ref{fig:4topxs} also illustrates the dependence of the total cross section on the HEFT coefficient $\abb$ for both truncation schemes. The light and dark blue bands represent projected 95\% confidence level sensitivities at the LHC~\cite{CMS:2023ftu} and HL-LHC~\cite{ATLAS:2025jdz}, respectively. 
The quadratic truncation yields tighter constraints, owing to the enhancement from squared terms at larger values of $\abb$. This contrasts with the Higgs signal strength fits discussed earlier, where both truncations gave nearly identical bounds, due to the stronger experimental constraints and the on-shell nature of the Higgs boson in those channels. The qualitative difference in four top quark final states is that the latter is rare, and the experimental sensitivity is comparably loose. Under such conditions, limits are susceptible to multiple BSM insertions, which can then also be used as a measure of uncertainty. In line with the lower experimental sensitivity to four top quark production, this theoretical uncertainty, which is also included in Fig.~\ref{fig:4topxs}, is large. The sensitivity to multiple operator insertions highlights the importance of accounting for non-linear effects in off-shell processes, such as four-top production. From the perspective of Monte Carlo event simulation, which is particularly relevant for the experimental collaborations, the linearisation of cross sections is also a non-trivial task for the present case of propagator modifications. 

For larger values of $\abb$ in Fig.~\ref{fig:4topxs}, the difference between the EoM-substituted and vanilla amplitudes becomes so pronounced that the theory error becomes greater than 50\%. The apparent saturation of this uncertainty at large 
$\abb$ is a numerical consequence of truncating the amplitude expansion at quadratic order and of the adopted uncertainty prescription, with the estimate being controlled by the highest retained powers of $\abb$ rather than signalling a physically stable regime. The error does not reflect a comparison between subsequent orders in $\abb$, which becomes large when the deviation from the SM is large. The latter is not reflected in the uncertainty associated with EoM vs. field redefinitions and should be considered as a separate and relevant quantity to establish perturbativity (we will briefly address this in the next section). Closer to the origin, a linear scaling in $\abb$ is still present but hidden beneath the sizeable uncertainties. Zooming in on small $\abb$ values reveals this linear regime, but for phenomenologically relevant values, the non-linear contributions dominate, highlighting the importance of accounting for higher-order effects in off-shell and rare processes such as four top quark production. 

\begin{figure}[!t]
    \centering
    \includegraphics[width=0.46\linewidth]{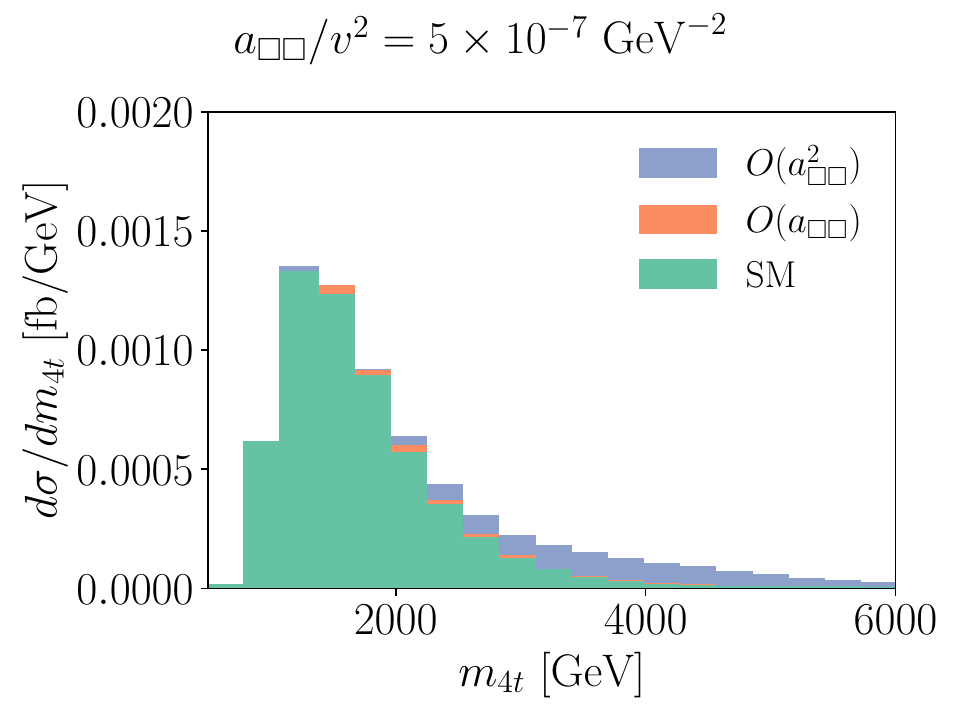}
    \includegraphics[width=0.46\linewidth]{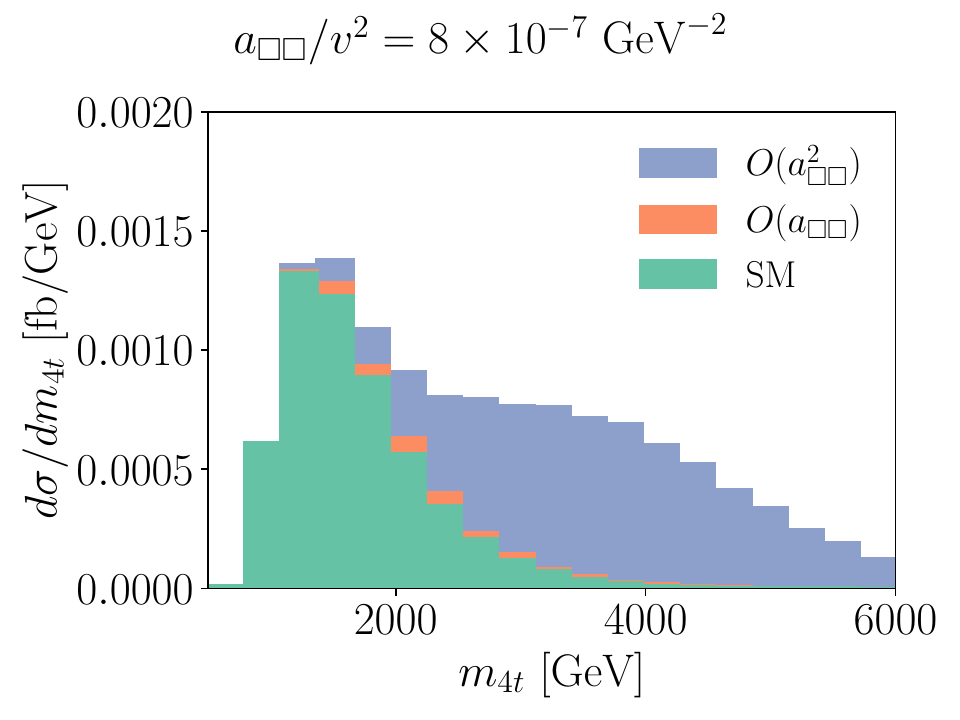}
    \caption{The four top invariant mass differential distributions for two benchmark points, including the SM, the linear and quadratic contributions from $\abb$.}
    \label{fig:minv4t}
\end{figure}

To establish context with the kinematic features on which these constraints are built, and to quantify relations of constraints with power counting and perturbative unitarity, Fig.~\ref{fig:minv4t} illustrates the four-top invariant mass ($m_{4t}$) distribution for two benchmark values of $\abb$. These benchmark points were chosen as a rough estimate of the exclusion limits on $\abb$ from the LHC, and the HL-LHC bands shown in Fig.~\ref{fig:4topxs}. At low masses and small values of $\abb$, both truncations behave similarly. However, for larger values of $\abb$, the high-mass tail exhibits pronounced deviations when the full quadratic structure is included. These enhancements are expected and arise from the growing contribution of higher-derivative interactions at large momentum transfers. Such kinematic distortions offer a complementary handle on effective operators beyond total rate measurements, reinforcing the role of four-top production as a precision probe of extended Higgs dynamics, e.g. in differential analyses beyond total rate measurements. This, of course, depends on the limit of the coefficient, which a priori may violate power-counting or unitarity constraints. These warrant independent consideration, and we turn to these questions in the next section.

\subsection{Notes on Perturbative Unitarity and Power Counting}
\label{sec:pcpu}
Theoretical consistency limits can be placed on the size of coefficients in any QFT, and these provide information complementary to theoretical uncertainty estimates. For completeness, we discuss them briefly here. We highlight that they provide, importantly, an upper bound rather than a quantitative assessment of error. We will specifically discuss constraints from power counting and perturbative unitarity.

\begin{figure}[!b]
    \centering
    \includegraphics[width=0.56\linewidth]{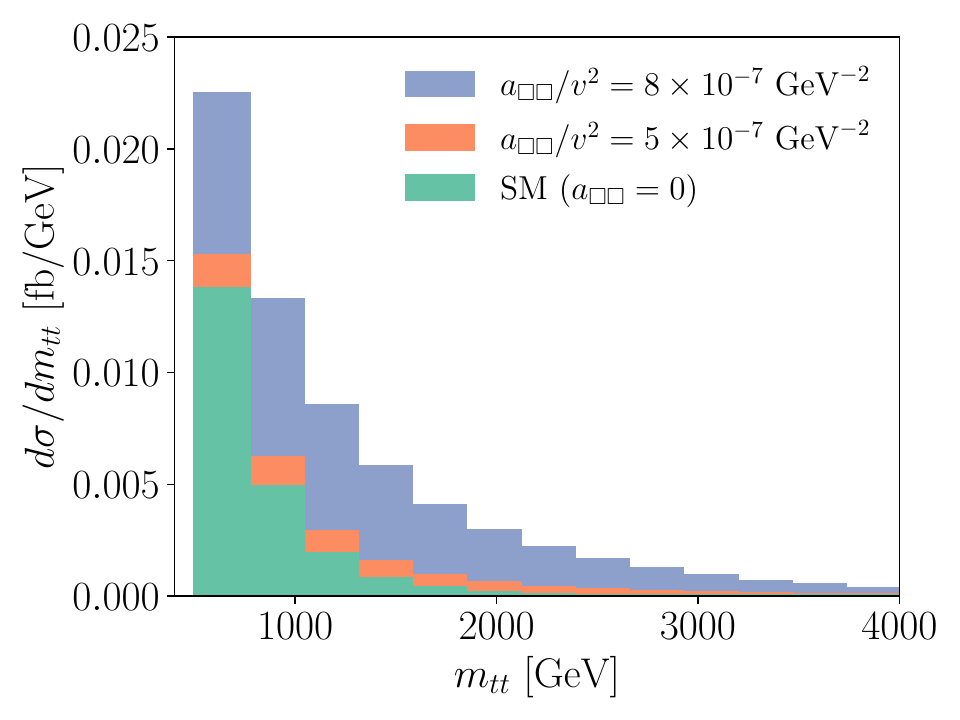}
    \caption{Invariant $t\bar{t}$ mass distribution for $pp \to t \bar{t} t \bar{t}$ at $\sqrt{s}=13~\text{TeV}$ (including quadratic contributions from $\abb$). \label{fig:minvtt}}
\end{figure}

\subsubsection*{Perturbative Unitarity} 
Perturbative unitarity~\cite{Logan:2022uus,Jacob:1959at,DiLuzio:2016sur} provides a key consistency check for effective field theories, ensuring that scattering amplitudes respect $S$-matrix unitarity at high energies. Any experimental analysis relies on a perturbative workflow, which is only self-consistent if it eventually yields a perturbative constraint. For higher-derivative operators such as $\obb$, which generate amplitudes that grow with energy, this requirement sets an upper limit on the energy scale where the effective description remains valid. Using the partial-wave arguments outlined in App.~\ref{app:unitarity}, we find that for our benchmark HEFT coefficient $ \abb/v^2 = 5\times10^{-7}~\text{GeV}^{-2} $, the theory remains perturbative up to $\sqrt{s} \lesssim 4.2~\text{TeV}$. Increasing the coefficient to $\abb/v^2 = 8\times10^{-7}~\text{GeV}^{-2}$ lowers the unitarity bound to $\sqrt{s} \lesssim 3.3~\text{TeV}$. 

These scales define the technical validity range of the effective~theory; the coefficient needs to be small enough so that the scales probed in the experiment stay within the range of validity.\footnote{Alternative approaches, such as `data clipping', have been proposed in the literature. We consider the deliberate exclusion of measured data for the sake of theoretical interpretability to be an unjustifiable practice.} To ensure that our predictions remain within this unitarity-safe regime, we examine the invariant mass distribution of the $t\bar{t}$ system reconstructed from the four-top final state. Since the off-shell Higgs decays into $t\bar{t}$, this invariant mass effectively traces the internal Higgs momentum in the dominant diagrams. As shown in Fig.~\ref{fig:minvtt}, for the $\abb$ values considered, the bulk of the events lies well below the corresponding unitarity bounds, indicating a high probability that the EFT remains perturbatively valid with a unitary tree-level $S$-matrix. The maximally probed energy scale, however, requires confirmation and dissemination by the experiments.

\subsubsection*{Power Counting}
We note that when applied to HEFT, $\mathbf{V}_\mu \equiv ( D_\mu U) U^\dagger \sim \partial$ is a non-linear field with a known expansion in $\pi/v$, yet the generic Higgs-dependent functions are assumed to depend on $h/f$ where $v\neq f$. The scale of new physics, meaning the mass scale of new resonances, is $\Lambda$, and in the application of the power counting formula to our case study. Further details are given in the appendix~\ref{sec:pc}, in particular Tab.~\ref{tab:HEFT_NDA} there. 
Following the NDA convention, we have
\begin{align}
    \frac{\abb}{v^2}=\frac{\hat a_{\Box\Box}}{\Lambda^2}=\frac{\hat a_{\Box\Box}}{(4\pi f)^2}\,.
\end{align}
A given choice of $\abb/v^2$ gives $\Lambda^2=\hat a_{\Box\Box}(v^2/\abb)$ and given $\hat a_{\Box\Box}\lesssim 1$ we obtain an upper bound on the cut-off for perturbative UV dynamics. The values chosen in Fig.~\ref{fig:minvtt}, this returns $\Lambda\lesssim O(\text{TeV})$ in line with the estimate from unitarity bounds. Note that, as opposed to the unitarity bound, the `$\sim$' here follows from the order one factors that are obtained from explicit computation of loop corrections that power-counting cannot predict.

\section{Conclusions}
\label{sec:conc}
Tracing the importance of field redefinitions is a standard approach in perturbative collider phenomenology for quantifying theoretical uncertainties arising from neglecting higher-order contributions. This is not a rigorous measure, but it enjoys broad community support. 

When going beyond renormalisable interactions, as in effective field theory approaches such as SMEFT or HEFT, redundant field redefinitions are necessarily intertwined with the effective operator expansion, which typically also introduces new momentum dependencies into scattering processes. The uncertainty in the EFT parameter interpretation crucially depends on the accuracy of the measurement, just like in any perturbative Quantum Field Theory. But non-linear momentum dependencies, e.g., in the Higgs sector, can inflate these uncertainties well above the naive expectations from renormalisable field theories.

In this work, we have systematically surveyed the uncertainties arising from truncating the EFT expansion of the HEFT Lagrangian. The starting point for our assessment of uncertainties is the difference in the use of equations of motion and field redefinitions. Only the latter does lead to the same theory and can be used without loss of generality~\cite{Criado:2018sdb}. While they do agree at leading order in the EFT expansion (in fact, a reduction of an operator basis is most commonly achieved with equations of motion), their difference at higher orders is the source of our proposed theory error.

Focusing on non-linear momentum dependencies in the Higgs sector, which are tell-tale signatures of HEFT, we provide numerical estimates of the uncertainty of the HEFT interpretation of data, highlighting different phenomenological circumstances. On the one hand, for inclusive Higgs observables, the anticipated HL-LHC precision is sufficient for the HEFT truncation error to play a subdominant role. On the other hand, for processes that fingerprint Higgs momentum dependencies through off-shell effects in rare final states, such as the production of four top quarks, the combined impact of momentum-dependent field redefinitions and projected measurement accuracy yields sizeable interpretation uncertainties. 

For our case study, unitarity bounds and power-counting arguments provided an estimate for an upper bound on coefficients which overlapped order-of-magnitude-wise with the parameter region in which the theory error grows close to $O(1)$. In this instance, these theory considerations could have been equivalently used to estimate the breakdown of the expansion. However, what our method does beyond these is to provide a quantitative error, defined over the entire perturbativity range, which is readily comparable to experimental uncertainties.

\section*{Acknowledgements}
We thank Matthew McCullough for helpful discussions.
R.A and S.U.R. are supported by STFC under Grant No. ST/X003167/1.
C.E. is supported by the IPPP Associateship Scheme. 
W.N. acknowledges support by the Deutsche Forschungsgemeinschaft (DFG, German Research Foundation) under Germany’s Excellence Strategy - EXC 2121 ``Quantum Universe" - 390833306. This work has been partially funded by the Deutsche Forschungsgemeinschaft (DFG, German Research Foundation) - 491245950.
\appendix
\section{Use of the EoM twice as two field redefinitions}
\label{sec:matching}
The use of the EoM twice in a given operator can be cast into two subsequent field redefinitions. One obtains, with superscripts marking order in $\abb$ and $\delta h_{1},\delta h_{2}\sim  O(\abb)$ and $x,y$ labelling spacetime and summed with Einstein's convention
\begin{align}
    S=&S^{(0)}+S^{(1)}\,,\\
    S'=& S^{(0)}+S^{(1)}+\delta h_{1,x} \delta S^{(0)}_x+\delta  h_{1,x} \delta S^{(1)}_x+\frac12 \delta h_{1,x} \delta^2 S^{0}_{xy}\delta h_{1,y}\,,\\
    S''=&S^{(0)}+S^{(1)}+(\delta h_{1+2,x}) \delta S^{(0)}_x+\delta  h_{1+2,x} \delta S^{(1)}_x+\frac12 \delta h_{1+2,x} \delta^2 S^{0}_{xy}\delta h_{1+2,x}\nonumber\\ & \hspace{9cm}+\delta h_{2,x}\delta S^{(0)}_y\delta^2 h_{1,xy}\,,\label{eq:FRtwice}
\end{align}
where $\delta h_{1+2}=\delta h_{1}+\delta h_{2}$. The last term is proportional to the EoM and can be removed with a higher order transformation $\delta h_3=-\delta h_2\delta^2h_1\sim {O}(\abb^2)$ while only modifying order $\abb^3$ terms. The error, or difference between EoM-application and field redefinition, takes again the form of Eq.~\eqref{eq:2nd_variation_err} with $\delta h=\delta h_1+\delta h_2$. Finally, we note that one could have used the EoM twice in $\obb$ with
\begin{align}
    \delta h_1&=-\frac{\abb}{v^2}\Box h, & \delta h_2&=\frac{\abb}{v^2}\left(m_h^2 h-\frac{\partial \lag_{\text{int}}^h}{\partial h} \right),
\end{align}
to obtain at the linear level in $\abb$ a term in the Lagrangian
\begin{align}
    -\frac{\abb}{v^2}\left(m_h^2 h-\frac{\partial \lag_{\text{int}}^h}{\partial h} \right)^2 .
\end{align}
Keeping the $O( \abb^2)$ terms in the field-redefinition of Eq.~\eqref{eq:FRtwice} does lead to the same predictions as the original theory, e.g. the amplitude of Eq.~\eqref{eq:vanfactors}.

\section{Perturbative Unitarity}
\label{app:unitarity}
\begin{figure}[!t]
\centering
    \parbox{0.3\textwidth}{\centering\begin{tikzpicture}
        \begin{feynman}
            \vertex (a) at (0, 0);
            \vertex (b) at (0, 1.5);
            \vertex (t1) at (-1.5, 2) {\(\large t\)};;
            \vertex (t2) at (1.5, 2) {\(\large t\)};;
            \vertex (W1) at (-1.5, -0.5) {\(\large W^\pm/Z\)};
            \vertex (W2) at (1.5, -0.5) {\(\large W^\pm/Z\)};
            
            \diagram* {
                (a) -- [scalar, edge label = \(\large h\)] (b),
                (t1) -- [fermion] (b) -- [fermion] (t2),
                (W1) -- [boson] (a) -- [boson] (W2),
            };
        \end{feynman}
    \end{tikzpicture}}
    \hspace{1cm}
    \parbox{0.5\textwidth}{
    \caption{Feynman diagram topologies representing $t$-channel Higgs-mediated $t~W^\pm/Z \to t~W^\pm/Z$ scattering. Not shown are non-Higgs electroweak Feynman diagram topologies that contribute to the scattering amplitude at the considered order. 
    \label{fig:twztwz}}}
\end{figure}
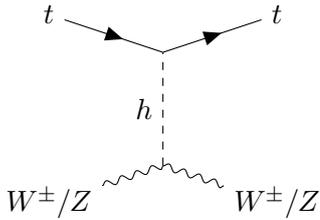

To determine the unitarity-violating scale in four top production associated with $\obb$, we consider high-energy $2 \to 2$ scattering processes that are directly sensitive to the momentum-dependent modifications introduced by this operator in this process. We focus on the processes $tW \to tW$, and $tZ \to tZ$, which receive contributions from $t$-channel diagrams involving an off-shell Higgs exchanged between fermion and gauge boson lines (cf. Fig.~\ref{fig:twztwz}).

The standard approach involves computing the helicity amplitudes $\mathcal{M}_{\lambda_1 \lambda_2 \to \lambda_3 \lambda_4}(s,\theta)$, where $\lambda_i$ denote the helicities of the external particles in the initial and final states. These amplitudes are decomposed into partial waves. The partial wave coefficients for total angular momentum $J$ and helicity difference $\lambda = \lambda_1 - \lambda_2$, $\lambda' = \lambda_3 - \lambda_4$, in the so-called Jacob-Wick formalism~\cite{Jacob:1959at} are given by
\begin{equation}
    a^J_{\lambda \lambda'}(s) = \frac{1}{32\pi s} \lambda_K^{1/4}(s,m_3^2,m_4^2)\lambda_K^{1/4}(s,m_1^2,m_2^2)  \int_{-1}^{1} {\text{d}}\hskip -0.05cm\cos\theta \, d^J_{\lambda \lambda^\prime}(\theta) \, \mathcal{M}_{\lambda_1\lambda_2 \to \lambda_3\lambda_4}(s,\theta),
\end{equation}
where $\lambda_K(\alpha,\beta,\gamma) = \alpha^2+\beta^2+\gamma^2 - 2 \alpha \beta - 2 \beta\gamma -2\gamma\alpha$ is the K\"all\'en-$\lambda$ function, and $d^J_{\lambda \lambda^\prime}(\theta)$ is the Wigner-$d$ function.

For the processes $tW^\pm \to tW^\pm$ and $tZ \to tZ$, the dominant contributions at high energies arise from the scattering of a top quark and a longitudinally polarised gauge boson~\cite{Chanowitz:1978mv}. We focus on a representative helicity configuration where the gauge boson is longitudinal (helicity $0$) and the top quark carries helicity $+\tfrac{1}{2}$. This corresponds to helicity differences $\lambda = \lambda' = +\tfrac{1}{2}$. In this case, the leading contribution to the partial wave amplitude stems from $J = \tfrac{1}{2}$. To derive the unitarity bounds, we numerically evaluate the amplitude $\mathcal{M}(s,\theta)$ for the relevant helicity configuration, project onto the $J = \tfrac{1}{2}$ partial wave using the expression above, and enforce the unitarity condition
\begin{equation}
\label{eq:unit}
    \left| \text{Re} \, a^{1/2}_{1/2,1/2}(s) \right| < \frac{1}{2}.
\end{equation}
This requirement ensures that the effective theory remains perturbative and predictive up to a given centre-of-mass energy $\sqrt{s}$. To enable a self-consistent comparison with experimental data, the maximum energy scale probed by a given measurement should respect the criterion of Eq.~\eqref{eq:unit}, ensuring that radiative corrections do not strongly distort the resulting limits.


\section{Power Counting}
\label{sec:pc}
\begin{table}[!t]
    \centering
    \setlength{\arrayrulewidth}{0.5mm}
    \setlength{\tabcolsep}{18pt}
    \renewcommand{\arraystretch}{1.6}
    \begin{tabular}{cccc}
        \hline
        Operator & mass dim & $d_\chi$ & NDA normalisation
        \\
        \hline
        $\mathbf{V}^2$ & 2 & 2 & $\frac{\Lambda^2}{(4\pi)^2} \mathbf{V}^2$ or $v^2 \mathbf{V}^2$ \\
        $(\partial h)^2$ & 4 & 2 &  $(\partial h)^2$\\
        $ m_h^2 h^2$ & 4 & 2 &  $m_h^2 h^2$\\
        $\mathbf{V}^4$ & 4 & 4 & $\frac{1}{(4\pi)^2} \mathbf{V}^4$ \\
        $(\square h)^2$ & 6 & 4 & $\frac{1}{\Lambda^2} (\square h)^2$ \\
        \hline
    \end{tabular} 
    \caption{HEFT operators and their power counting. The rightmost column represents the appropriate normalisation factors from NDA. The column labelled $d_\chi$ displays the chiral dimension for the operators.}
    \label{tab:HEFT_NDA}
\end{table}
Graph-theory results can be applied to the loop expansion to count factors of $4\pi$ for an arbitrary diagram. Combining this with dimensional analysis, one arrives at an estimate for the contribution of an operator to any other operator, up to order one coefficients\footnote{These could, however, be sizeably smaller than order one, with notable reasons for it being symmetry or holomorphy/helicity sum rules.}. This result suggests a normalisation for coefficients $a \to \hat a$ such that $\hat a\sim 1$ signals the breakdown of the loop expansion~\cite{Manohar:1983md}. To be precise, with this normalisation an operator $1$ with coefficient $\hat a_1$ would produce a loop contribution to another operator $2$ as $\hat a_{2,\text{eff}}\sim \hat a_2+O (1)\times \hat a_1$ and the same for $\hat a_{1,\text{eff}}$. The limit $\hat a\sim 1$ is therefore the limit of strong dynamics. This, however, does not mean that one has to give up the EFT, in fact, the most prominent use of power counting is the EFT of chiral perturbation theory, e.g.~\cite{Weinberg:1968de,Gasser:1984gg,Weinberg:1991um,Leutwyler:1996qg}.

When applied to HEFT, this procedure is not as straightforward to implement and has led to some discussion~\cite{Buchalla:2013eza,Gavela:2016bzc,Buchalla:2016sop}. Here let us simply borrow the formula of~\cite{Gavela:2016bzc} for comparison.
The NDA master formula in $d$ dimensions for an arbitrary EFT operator involving derivative $(\partial)$, scalar $(\phi)$, gauge boson $(A)$, and fermion $(\psi)$ fields and gauge $(g)$, Yukawa $(y)$, and quartic scalar $(\lambda)$ coupling constants can be written as
\begin{multline} 
\label{eq:nda_master_d}
    \frac{\Lambda^d}{(4\pi)^{d/2}} \left[\frac{\partial}{\Lambda}\right]^{N_p}\,
    \left[\frac{(4\pi)^{d/4} \phi}{\Lambda^{(d-2)/2}}\right]^{N_\phi} \,
    \left[\frac{(4\pi)^{d/4} A}{\Lambda^{(d-2)/2}}\right]^{N_A} \,
    \left[\frac{(4\pi)^{d/4} \psi}{\Lambda^{(d-1)/2}}\right]^{N_\psi} \, \\
    \left[\frac{g}{(4\pi)^{d/4}\Lambda^{(4-d)/2}}\right]^{N_g} \,
    \left[\frac{y}{(4\pi)^{d/4}\Lambda^{(4-d)/2}}\right]^{N_y} \,
    \left[\frac{\lambda}{(4\pi)^{d/2}\Lambda^{(4-d)}}\right]^{N_\lambda} \,,
\end{multline}
where
\begin{align}
    \Lambda = (4 \pi)^{\frac{d}{2(d-2)}} f\,.
\end{align}
This further generalises if we include the mass term $(m_i)$ or the trilinear coupling of scalar~$(\text{k})$:
\begin{align}\label{eq:nda_mass}
    \left[\frac{m_\phi^2}{\Lambda^2}\right]^{N_{m_\phi}}\left[\frac{m_\psi}{\Lambda}\right]^{N_{m_\psi}} \left[\frac{\text{k}}{(4\pi)^{d/4}\Lambda^{(6-d)/2}}\right]^{N_{m_\phi}}\,.
\end{align}
The combination of Eqs.~\eqref{eq:nda_mass} and \eqref{eq:nda_master_d} provides the complete NDA master formula.


\bibliographystyle{JHEP}
\bibliography{references}
	
\end{document}